\documentclass[acus]{JAC2000}

\usepackage{graphicx}

\begin{document}
\title{\flushright{WEAP023}\\[15pt] \centering Modernising the ESRF control system 
with GNU/Linux}

\author{A.G\"otz, A.Homs, B.Regad, M.Perez, P.M\"akij\"arvi, W-D.Klotz\\
ESRF, 6 rue Jules Horowitz, Grenoble 38043, FRANCE}

\maketitle

\begin{abstract}
The ESRF control system is in the process of being modernised. The present control
system is based on VME, 10 MHz Ethernet, OS9, Solaris, HP-UX, NFS/RPC,
Motif and C. 
The new control system will be based on compact PCI, 100 MHz Ethernet,
Linux, Windows, Solaris, CORBA/IIOP, C++, Java and Python. The main frontend
operating system will be GNU/Linux running on Intel/x86 and Motorola/68k. 
Linux will also be used on handheld devices for mobile control. 
This poster describes how GNU/Linux is being used to modernise the control 
system and what problems have been encountered so far\footnote{work supported by J.Klora, J.M.Chaize and P.Fajardo}.

\end{abstract}

\section{Introduction}
The ESRF control systems control 3 accelerators and 32 beamlines.
They have been built using the same technology and are completely compatible.
They were built 10 years ago based on the state-of-the-art technology ten
years ago.
This included VME, 10 MHz Ethernet, OS-9, Solaris, HP-UX, NFS/RPC, Motif
and C. Most of these technologies have not evolved over the last few
years.
In our search for better tools, support, ease of programming, and overall
stability and quality we have put all our old technologies to the test.
Our main criterium was which technology or tool will allow us to offer
users a better control system.
A better control system means one which offers more features  to users
without losing any of the present good features. 

The result of this technology survey was 100 MHz Ethernet, VME (for the 
existing hardware), CompactPCI (cPCI) and PCI for new hardware,
Linux as main frontend operating system, Windows for commercially supported 
hardware and software, Solaris and GNU/Linux as main desktop operating systems,
CORBA/IIOP as new network protocol, C++, Java and Python as main programming
languages.


\section{Why GNU/Linux ?}

What does GNU/Linux offer that other systems doesn't offer ? 
\begin{enumerate}
\item
FREEDOM ! Freedom in this context means access to all the source code
so that it can be compiled, understood and improved. 
An additonal freedom is the freedom from supplier pressure and fees. 
\item
Technology we know well (Unix) and
which is conceptually simple to understand and program. 
This is an important feature in our case because we need to develop device 
drivers. In addition to being easy to understand it is well-documented.
\item
A rich set of software packages. Almost all known
sourceware packages have been developed or ported to GNU/Linux.
\item
It is easy to manage in a network environment and has
excellent support for all network protocols. Because our control
systems are distributed over the network this played a strong role in
our choice for GNU/Linux.
\end{enumerate}

\section{Linux/m68k + VME}
The ESRF has over 200 VME crates installed.
This represents an investment of millions of Euros as well as many 
tens of years of work hardware and software development.
Any modernization project must take this investment into account. 
The modernization foresees two ways to do this - using the Motorola CPU's (MVME-162)
to run GNU/Linux directly or replacing the CPU with a bus extender which allows
the VME bus to be controlled from PC running Linux/x86. This section describes
the first option. The bus extender solution is discussed in the next section.

For Linux/m68k we use the Debian distribution 2.1. 
It can be downloaded from the Debian website\footnote{http://www.debian.org} and 
is available in source code and binary format.
The standard kernel (we are running kernel 2.2.10) includes the support for the 
Motorola CPU port (originally done by Richard Hirst\footnote{rhirst@sleepie.demon.co.uk}).
We run all our Linux/m68k crates without harddisk (diskless). The root disk
is NFS mounted readonly. In addition there is a RAM disk for /etc, /dev, /var and /tmp.
This means crates can be switched on/off without risk of losing data nor do we have
to do fsck's.
We have rewritten device drivers for all our main VME cards.
For many of them we subcontracted the driver writing for the first version
to Richard Hirst (later Linuxcare). 
Maintenance and further development is now done in house.
Client programs communicate with the hardware via the network using TACO/TANGO 
device servers (cf. below).
We use the GNU tools for compiling and debugging (g++ and gdb).

Our experience with Linux/m68k compared to OS-9 (the commercial operating system
we were using previously) is that it is at least if not more stable, 
the TCP/IP implementation is more efficient and robust and that it is 
easy to add new features to our software using standard techniques like 
multithreading.

\section{Linux/x86 + bus extenders}

The modernization project of the instrument control at the ESRF using GNU/Linux 
supports two main hardware platforms: PCI/cPCI and VME. 
The former provides access to the most recent interface boards developed 
for a highly demanding market, and hence, with better performance/price ratios. 
The latter is needed for a gradual transition between the current VME 
instrumentation and the PCI technology. 
VME boards can be controlled from a Motorola MVME CPU or from a PC 
through a PCI/VME bus extender, both running GNU/Linux as OS. 

As it was said before, the modernization project also includes the cPCI 
platform, which, in combination of PCI/cPCI bus extenders, notably increases
the flexibility in the hardware configuration. 
It is well known that due to the dynamic resource configuration in the 
PCI specification, identical boards are only distinguishable by their 
slot position in the bus. 
Most of the drivers available for PCI boards enumerate the boards in the 
same order they are found by the BIOS / OS at boot time. 
This means that the board identification number will change when a 
similar one, situated before in the PCI bus structure, is removed. 
Moreover, if we apply the same logic to slave cPCI crates, their 
bus numbers will change when another is removed. 
To solve this problem a differentiation between physical numbers, 
those used by the drivers, and logical numbers used by applications is made. 
The mechanism responsible to make this mapping keeps track of the boards 
present in the system and detects any change in the bus configuration. 
Any non-trivial change is informed to the user, avoiding wrong addressing 
to the boards. 
The position of the boards are presented to the user in terms of chassis 
and slot, which are translated to PCI device numbers by hardware specific 
mappings.

Setups based on the PCI architecture have been mounted using both a 
desktop PC and an industrial PC that implements the PICMG standard. 
Remote VME crates are controlled through SBS Technologies PCI/VME bus 
extenders, and cPCI crates are directly linked to the main PCI bus by 
means of National Instruments MXI-3 PCI/cPCI/PXI bus adapters. 
These adapters expand by a large factor the amount of hardware that 
can be managed by a single host. Furthermore, both MVME and PCI GNU/Linux 
can independently control boards in the same crate, providing even 
more possibilities for the VME - PCI transition. 


\section{Device Drivers}
In order to use the same device driver codes in both systems, an interface 
layer was implemented to manage I/O addresses and IRQs, taken from the 
module parameters at load time. 
In the bus coupler configuration, this interface does the necessary PCI to 
VME address mappings during the initialization of the VME board drivers, 
allowing boards on different   (remote) VME buses to be controlled from 
the same host as local. 
This interface also exports automatically the state and configuration of 
each board to the {\tt /proc} virtual file system. 

In experiment automation it is often very useful to record the value of 
several magnitudes when an event occurs. 
Such an event can be generated by a hardware signal or by a software condition. 
To provide this functionality a buffering mechanism was developed in the 
kernel, named hook after a similar facility developed at the ESRF for OS/9 
drivers, which hooks data on hardware interrupts. 
The values to be written in the buffer are run-time configurable 
by specifying the driver name, the board and the channel to be read. 
Each driver that can export its channels will register with the hook 
module during initialization. 
When an application wants to read one of its channels, the hook asks 
for the necessary actions to be done. 
If the actions are just simple register read/write operations 
(one single read is very common), they are returned in the form of a "program". 
Otherwise, if the process is more complicated, a pointer to a function is saved.
One source of hook events is a timer provided by the hook itself, 
which attaches to the system software timer, and hence has a minimum 
repetition period of 10 ms on standard installations. 
Higher rates can be achieved with hardware interrupts generated by 
counter/timer boards like the ESRF VCT6.

Not all the boards allow a fast reading of their registers, and 
the system should not wait in an interrupt handler (actually a bottom half handler). 
This problem can be overcome with an asynchronous buffer writing, as long as it is 
done before the next event arrives. 
Finally, the hook buffer can be filled in linear mode, which stops acquisitions 
when the end of the buffer is reached, or in circular mode for continuous measurement.

\section{Device Servers}

The device drivers are the first layer in our control system architecture.
The second layer is called the {\em device server} layer and provides
transparent network access. 
This means hardware can be shared transparently between geographically
separated parts of the accelerator and/or beamline, thereby adding a
layer of flexibility which would otherwise not be available (except by
recabling).
The device servers at the ESRF are of two flavours. The original flavour
called TACO\footnote{http://www.esrf.fr/taco} uses the ONC/RPC as network
protocol and is a lightweight protocol. It has the advantage that the
ONC/RPC runs everywhere where NFS runs.
The second flavour called TANGO\footnote{http://www.esrf.fr/tango}
is based on CORBA and uses the IIOP protocol for the network layer.
CORBA is slightly more heavyweight compared to ONC/RPC but offers
more high-level services.
Both flavours of device servers offer synchronous, asynchronous and 
event-driven communication paradigms and a database for permanent storage.
A large number (hundreds) of device servers have been written at
the ESRF and other sites (FRM II, Lure, HartRAO). Refer to the
websites for more information.

\section{Administration}

The challenge we are facing with the modernization of the control system is 
not only to be able to provide the best combination of 
operating system+hardware, but also to be able to do the system 
administration  of the system installed all over the site. 
Administration means two important things :
\begin{itemize}
\item
quick recovery of a  system after a failure.
\item
new release of the system.
\end{itemize}

Our present control system is based on VME / OS9 diskless systems. 
These OS9 systems are served by bootp servers which give them an 
identity and then  downloads the kernel onto the VME crate at startup.
The VME crates then mounts all the same remote file system using NFS.
Thus,  for example  if a CPU fails we have just to change it and press 
the ON button.
This type of action takes less than 15 minutes.

With the modernized control system,  BOOTP technique is 
replaced with DHCP which is based on BOOTP but has more powerful features. 
In this way DHCP allows for dynamic allocation of the network address. 

The compact PCI crates that are being installed are equipped with  a hard disk. 
It is inconceivable to think that numerous systems that will be running in the 
future have each their  own configuration. 
In this case it should be almost impossible to do the system administration  
and to provide a good service to the users of these systems.

The decision was taken to do a base system and to duplicate it as many time 
as it will be needed. 
This technique is also called  "cloning".
The ESRF has bought a commercial product called REMBO (for REMote BOot). 
This tool is able to deal with DHCP technology to give a network identity 
to a client that broadcasts a DHCP request.
Moreover this tool can be  used to make a base image of any system 
(Linux / Windows) and to upload that image on a computer on which  a 
Rembo server has been installed .
REMBO is a cloning tool and a backup utility as well. The administration of the Compact 
PCI can be improved with the capability of REMBO to do differential image.  
We use this feature to manage different hardware configuration starting from 
a base system. 
For example the crates dedicated to vacuum and the crates dedicated to the 
front end system have the same base hardware (same crate, same CPU , 
network card) and therefore will have the same base image (same 
operating system, network drivers, etc.).
But since these crates do not have the same I/O boards installed
a differential image for each type of hardware configuration  will be made. 
In case of failure (crash of the hard disk) the tool will be able to 
rebuild the  whole system from the base image and the differential one .

\section{Handheld devices}

We have long had the need for handheld portable controllers for motors,
and other hardware which need local tuning far from the control room.
Enter the new generation of handheld devices, wireless Ethernet and
Linux. We have used the iPAQ from Compaq running GNU/Linux and an X11
based client (Labview for example) to control motors remotely
using a simple one-click interface.
The fact of choosing GNU/Linux makes the task of network and graphical
display much easier than if we were using Windows CE for example.

\section{Realtime}
{\em ``Linux isn't realtime''.} 
This is true therefore we do not claim to do any realtime with GNU/Linux.
However we have found that for our applications we need little or no
realtime. Most realtime needs are delegated to hardware or DSP's.
Where we do need soft realtime (i.e. 99\% guarantee) we use interrupt
routines in device drivers. We measured an interrupt response time
of $<50 \mu s$ for Linux on 68k and x86. 
Using a driver interrupt routine we achieve a 
soft realtime response of $500 \mu s$ for a function generator 
(providing we do not recompile the kernel at the same time !).
For the rest of our applications {\em ``as-fast-as-possible''} is good enough.
And for that GNU/Linux on commodity hardware is surprisingly good.


\section{Problems}
GNU/Linux is not without problems. 
The main problems we have identified so far are :
\begin{itemize}
\item
the standard GNU/Linux distributions are not easily adapted to running on
diskless systems
\item
most commercial hardware does not have GNU/Linux drivers but Windows drivers
\end{itemize}

\section{Conclusion}
There is a viable alternative to Windows 95/98/ME, NT/2000/CE/XP for 
building control systems and it is called GNU/Linux !
Linux is sufficiently mature for the task and even offers some advantages
i.e. it is easier to program, is better adapted to distributed control
and is free of commercial pressure.

%
%

\end{document}